\documentclass[12pt]{article}

\usepackage{amsfonts,amsmath,amssymb}
\usepackage{bm}
\usepackage{cite}

\allowdisplaybreaks[1]

\newcommand{\R}{\mathbb{R}}
\newcommand{\Z}{\mathbb{Z}}

\begin{document}

\begin{titlepage}

\begin{flushright}
arXiv:0812.4883 [hep-th]\\
KUNS-2176\\
YITP-08-99
\end{flushright}

\vspace{0.1cm}

\begin{center}
  {\LARGE
The Kerr/CFT Correspondence and String Theory  
  }
\end{center}

\vspace{0.1cm}

\begin{center}
         Tatsuo A{\sc zeyanagi}$^{\dagger}$\footnote
           {
E-mail address : aze@gauge.scphys.kyoto-u.ac.jp}, 
         Noriaki O{\sc gawa}$^{\S}$\footnote
         {
E-mail address : noriaki@yukawa.kyoto-u.ac.jp} and    
         Seiji T{\sc erashima}$^{\S}$\footnote
           {
E-mail address : terasima@yukawa.kyoto-u.ac.jp}

\vspace{0.3cm}

${}^{\dagger}$
{\it Department of Physics, Kyoto University,\\
Kyoto 606-8502, Japan}\\

${}^{\S}$
{\it Yukawa Institute for Theoretical Physics, Kyoto University,\\
     Kyoto 606-8502, Japan}
\end{center}

\vspace{1.5cm}

\begin{abstract}
The Kerr/CFT correspondence is a holographic duality
between 
a two dimensional chiral conformal field theory (CFT) and
the very near horizon limit of an extremal black hole,
which includes an AdS$_2$ structure.
To understand the dual chiral CFT$_2$,
we apply the Kerr/CFT correspondence 
to a certain class of black holes embedded in string theory,
which include the D1-D5-P and the BMPV black holes,
and obtain the correct entropies for the black holes microscopically.
These have an AdS$_3$ structure in the near horizon geometry 
and an AdS$_2$ structure in the very near horizon geometry.
We identified one of the two Virasoro symmetries in the nonchiral CFT$_2$
dual to the AdS$_3$, i.e. in the AdS$_3$/CFT$_2$,
with the Virasoro symmetry in the chiral CFT$_2$ dual to the AdS$_2$,
i.e. in the Kerr/CFT correspondence.
We also discuss a way to understand the chiral CFT$_2$ 
dual to generic extremal black holes.
A kind of universality for the very 
near horizon geometries of extremal black holes will be important
for the validity of the Kerr/CFT correspondence. 
Based on this analysis, we propose that
the Kerr/CFT correspondence
can be understood as a decoupling limit 
in which only the ground states remain.
\end{abstract}
\end{titlepage}

\tableofcontents

\section{Introduction}
\label{introduction}

Black holes are still a mystery and provide us
valuable information to understand a microscopic 
theory underlying Einstein's general relativity. 
Toward the formulation of the microscopic theory,
they are expected to play the same role
as hydrogen atoms did in the early stage of quantum mechanics.
In 1996, Strominger and Vafa succeeded in 
microscopically calculating the Bekenstein-Hawking entropy of 
a BPS black hole by using string theory \cite{Strominger:1996sh}.
This success is one of the most stringent pieces of evidence to believe 
that string theory or nonperturbative formulation of it 
is a promising candidate for quantum theory of gravity. 

Recently the Bekenstein-Hawking entropy of the 4D extremal 
Kerr black hole was calculated \cite{Guica:2008mu} 
by applying Brown-Henneaux's method \cite{Brown:1986nw}%
\footnote{
In \cite{Carlip:1998wz},
Brown-Henneaux's method was applied to black holes in a different way,
considering not the near horizon geometry but
the whole black hole geometry,
imposing a boundary condition around the horizon
and taking diffeomorphisms on the ($t,r$)-plane.
For related works to that approach, see
\cite{Solodukhin:1998tc,%
Carlip:1999cy,Park:2001zn,Carlip:2002be,%
Cadoni:1998sg,Cadoni:1999ja,Castro:2008ms}, for example.
}.
Roughly speaking,
the near horizon geometry of it
is written as a $U(1)$ fibrated AdS$_2$ geometry
and the Virasoro symmetry of the dual chiral CFT$_2$ is identified
with a class of diffeomorphisms which preserve an appropriate
boundary condition.
This analysis is
somehow different from the usual AdS/CFT correspondence
\cite{Maldacena:1997re,Gubser:1998bc,Witten:1998qj},
and an interesting point
is that the $U(1)$ part rather than the $SL(2,\R)$ part of the isometry
is enhanced to the Virasoro symmetry and it is not clear why
the enhancement occurs in such a manner.

After that, this method, the Kerr/conformal field theory (CFT) correspondence,
has been applied
to various black holes with such a fibration structure in the
near horizon geometry
\cite{Hotta:2008xt,Lu:2008jk,Azeyanagi:2008kb,Hartman:2008pb,Nakayama:2008kg,
Chow:2008dp,Isono:2008kx},
and
the correct entropy
is reproduced microscopically for each of the systems.
In particular,
as shown in \cite{Azeyanagi:2008kb, Hartman:2008pb},
the Kerr/CFT correspondence can be applied
to any $U(1)$ symmetry instead of the rotational symmetry of space
of an extremal black hole with a finite horizon area.
However,
the origin of this correspondence is not clear yet.
In \cite{Azeyanagi:2008kb},
there are two different boundary conditions
which give different Virasoro symmetries for one black hole geometry,
although both give the correct entropy.
This fact also makes the correspondence mysterious.
To understand the Kerr/CFT correspondence well,
it is
worthwhile to apply it to well-known black holes embedded
in string theory. This is the main topics of this paper.

In this paper, we apply the method to a certain class
of black holes which are constructed as a rotating 
D1-D5-P system in string theory \cite{Cvetic:1996xz,Cvetic:1998xh}. 
Especially
this class includes the BMPV black hole 
\cite{Breckenridge:1996is, Breckenridge:1996sn}
and the nonrotating 
D1-D5-P black hole
\cite{Strominger:1996sh,Callan:1996dv,Maldacena:1996ky},
which preserve supersymmetries.
This class of black holes contain an AdS$_3$ structure 
in the usual
``near horizon'' geometry, which is familiar
in the context of AdS/CFT correspondence.
Going into this AdS$_3$ throat corresponds to a decoupling limit,
where only the low energy states
compared with the string scale remain.
At the bottom of this AdS$_3$ throat,
we find a BTZ black hole,
where we can take a further near horizon limit
and find a $U(1)$ fibrated AdS$_2$ structure.
We call this second limit the ``very near horizon limit,''
according to \cite{Strominger:1998yg},
in order to distinguish it from the first.
We can apply the Kerr/CFT correspondence to this
``very near horizon geometry.''
This may imply that there is a further decoupling limit
for the nonchiral CFT$_2$ dual to the AdS$_3$.
We discuss that the very near horizon limit 
is the low energy limit to the lowest energy states,
namely the ground states
with the fixed asymptotic charges.%
\footnote{
Discussions relating AdS$_2$ limits to ground states
are also found in \cite{Strominger:1998yg,Maldacena:1998uz,Gupta:2008ki}.
}

From this hierarchical structure,
we identify
one of the two Virasoro symmetries in the nonchiral CFT$_2$ dual to the AdS$_3$
\cite{Strominger:1997eq},
i.e., in the AdS$_3$/CFT$_2$ correspondence,
with the Virasoro symmetry in the chiral CFT$_2$ dual to the AdS$_2$,
i.e., in the Kerr/CFT correspondence.
For preceding analysis to
relate AdS$_3$ and AdS$_2$,
see \cite{Maldacena:1998bw,Azeyanagi:2007bj,Castro:2008ms,Hotta:2008xt}.
We also obtain the correct entropies for this class of the black holes
by the Kerr/CFT correspondence.
Note that the representation of the Virasoro algebra
of the chiral CFT$_2$ includes states
in different very near horizon geometries
because the very near horizon geometries depend on
the momentum $P$.
Therefore, the whole chiral CFT$_2$ does not live
on the very near horizon geometry with a fixed $P$.
We need more states than those on the very near horizon geometry
to realize it.

We also discuss a way to understand the chiral CFT$_2$ for
more generic cases.
In the Kerr/CFT correspondence
the central charge for the Virasoro algebra in the rotating direction
is roughly proportional to the angular momentum $J$ of the black hole.
However, as stated above,
the chiral CFT$_2$ is not expected to live in
the very near horizon geometry.
It is natural to ask what is the origin of the chiral CFT$_2$.
As shown in
\cite{Kunduri:2007vf,Astefanesei:2007bf,Kunduri:2008rs,Kunduri:2008tk},
the very near horizon geometries of
extremal black holes are highly constrained and then
there will be a kind of universality for such geometries.
An example of such universality
is the near horizon extremal geometries of
the 5D slow-rotating Kaluza-Klein black hole
\cite{Gibbons:1985ac,Rasheed:1995zv,Matos:1996km,Larsen:1999pp} and
the 5D Myers-Perry black hole \cite{Myers:1986un}
on an orbifold $\R^4/\Z_{N_6}$,
which was shown in \cite{Kunduri:2008rs}.
The Kerr/CFT correspondence in the compactified direction
of the former geometry \cite{Azeyanagi:2008kb}
will account for the Kerr/CFT correspondence in the rotational
direction of the latter geometry \cite{Lu:2008jk}.
From this, we expect that the chiral CFT$_2$ of the Kerr/CFT correspondence
is originated in some high energy completion
of the very near horizon geometry,
for instance an AdS$_3$ throat structure.
This may also mean that
the same very near horizon geometry of different black holes
have corresponding different boundary conditions in
the Kerr/CFT correspondence.

Organization of this paper is as follows.
In section \S\ref{review_kerr/cft}, we review
the Kerr/CFT correspondence.
In section \S\ref{near_horizon},
we explain the rotating 5D black holes
and apply Brown-Henneaux's procedure
to the near horizon geometry.
In section \S\ref{very_near_horizon},
we take the very near horizon limit for the near horizon geometry,
carry out the Kerr/CFT procedure for it
and compare the results with those of \S\ref{near_horizon}.
In section \S\ref{decoupling_limit},
we try to understand the very near horizon limit
as a decoupling limit
and discuss why the Kerr/CFT correspondence works for
generic extremal black holes.
Finally, section \S\ref{conclusions} is devoted to conclusions.

\section{A Review of Kerr/CFT Correspondence} 
\label{review_kerr/cft}

In \cite{Guica:2008mu}, it is proposed that the
4D extremal Kerr black hole is dual to a chiral CFT$_2$.
The metric of a generic Kerr black hole is written as
\begin{align}
ds^2 = -\frac{\Delta}{\rho^2}
\Bigl(d\hat{t}-a\sin^2\theta d\hat{\phi}\Bigr)^2
&+\frac{\sin^2\theta}{\rho^2}\Bigl((\hat{r}^2+a^2)d\hat{\phi}-ad\hat{t}\Bigr)^2
+\frac{\rho^2}{\Delta}d\hat{r}^2+\rho^2d\theta^2,\\
\Delta=\hat{r}^2-2M\hat{r}+a^2,& \quad\qquad \rho^2=\hat{r}^2+a^2\cos^2\theta,
\\
a=\frac{G_4J}{M},&\quad\qquad M= G_{4}M_{ADM},
\end{align}
where $G_4$ is the 4D Newton constant and $J$, $M$ and $M_{ADM}$
are the angular momentum, the geometric mass,
and the Arnowitt-Deser-Misner(ADM) mass, respectively.
The event horizon is located at $r_+=M+\sqrt{M^2-a^2}$,
and the Bekenstein-Hawking (BH) entropy is
\begin{align}
S_{\mathit{BH}} = \frac{2\pi Mr_+}{G_4}.
\label{kerr_sbh}
\end{align}
The Hawking temperature and the angular velocity at the horizon are
\begin{align}
T_H = \frac{r_+-M}{4\pi M r_+}, \quad  \Omega_H=\frac{a}{2Mr_+}.
\end{align}

This black hole is extremal when
\begin{align}
M = \sqrt{G_4J}.
\label{kerr_extremal_condition}
\end{align}
By defining
\begin{align}
t=\frac{\epsilon \hat{t}}{2M},\quad y=\frac{\epsilon M}{\hat{r}-M},
\quad \phi = \hat{\phi}-\frac{\hat{t}}{2M},
\label{very_near_kerr}
\end{align}
and introducing the nonextremality parameter
\begin{align}
\delta=M-\sqrt{G_4J},
\label{kerr_nonextremality}
\end{align}
the near horizon extremal limit \cite{Bardeen:1999px} is
taken as
\begin{align}
\delta\to 0,\; \epsilon\to 0
\quad \text{while}\;\delta/\epsilon\to 0.
\label{kerr_near_horizon_extremal_limit}
\end{align}
Then the near horizon extremal Kerr geometry is written as
\cite{Bardeen:1999px}
\begin{align}
ds^2
&= 2G_4J\Omega^2 \biggl[
\frac{-dt^2+dy^2}{y^2} +d\theta^2
+ \Lambda^2 \Bigl(d\phi+\frac{dt}{y}\Bigr)^2\biggr],
\\
&\qquad \Omega^2 = \frac{1+\cos^2\theta}{2},\qquad
\Lambda =\frac{2\sin\theta}{1+\cos^2\theta},
\end{align}
or, by introducing a global coordinate system,
\begin{align}
y = \Bigl(\cos\tau\sqrt{1+r^2}+r\Bigr)^{-1},\quad
t = y\sin\tau\sqrt{1+r^2},\nonumber\\
\phi = \varphi
+\log\Bigl(\frac{\cos\tau+r\sin\tau}{1+\sin\tau\sqrt{1+r^2}}\Bigr),
\end{align}
as
\begin{align}
ds^2 = 2G_4 J\Omega^2
\Bigl[
-(1+r^2)d\tau^2+\frac{dr^2}{1+r^2}
+d\theta^2+\Lambda^2(d\varphi+rd\tau)^2
\Bigr].
\label{nhek}
\end{align}

The generators of the Virasoro symmetry of the chiral CFT$_2$ are
identified with those of a class of diffeomorphisms
which preserve an appropriate boundary condition on
the near horizon geometry.
Then it is found that the nontrivial part
of the diffeomorphisms, or the asymptotic symmetry group (ASG),
contains diffeomorphisms generated by
\begin{align}
\zeta_n=-e^{-in\varphi}\partial_{\varphi}-inre^{-in\varphi}\partial_r
\quad (n=0,\pm1,\pm2,\cdots),
\label{kerr_zeta}
\end{align}
and they generate a Virasoro algebra without a central charge
\begin{align}
i[\zeta_m, \zeta_n]_{\mathit{Lie}} =(m-n)\zeta_{m+n}.
\label{virasoro_zero}
\end{align}
Here we notice that $\{\zeta_{n}\}$ contains $\partial_{\varphi}$,
not $\partial_{\tau}$.

By following the covariant formalism of the ASG
\cite{Barnich:2001jy, Barnich:2007bf, Abbott:1981ff,Iyer:1994ys,%
Anderson:1996sc, Torre:1997cd, Barnich:1994db, Barnich:2000zw,%
Barnich:2003xg, Compere:2007az},
for the $d$-dimensional ($d\ge 3$) case in general,
the conserved charge $Q_{\zeta}$ associated with an element $\zeta$ is defined by
\begin{align}
Q_{\zeta} =\frac{1}{8\pi G_d}
\int _{\partial\Sigma} k_{\zeta}[h,\bar{g}],
\label{charge_zeta}
\end{align}
where $\partial\Sigma$ is a ($d-2$)-dimensional spatial surface at infinity,
$G_d$ is the Newton constant and
\begin{align}
k_{\zeta}[h,\bar{g}]
&=\frac{1}{2}
\Bigl[
\zeta_{\nu}D_{\mu}h -\zeta_{\nu}D^{\sigma}h_{\mu\sigma}+
\zeta^{\sigma}D_{\nu}h_{\mu\sigma} \nonumber\\
&\qquad +\frac{1}{2}h D_{\nu}\zeta_{\mu}
-h_{\nu\sigma}D^{\sigma}\zeta_{\mu}
+\frac{1}{2}h_{\nu\sigma}(D_{\mu}\zeta^{\sigma}+D^{\sigma}\zeta_{\mu})
\Bigr]
*_d\!(dx^{\mu}\wedge dx^{\nu}).
\label{current_zeta}
\end{align}
Here $*_d$ represents the Hodge dual,
$\bar{g}_{\mu\nu}$ is the metric of the background geometry
and $h_{\mu\nu}$ is a deviation from it.
We also notice that the covariant derivative
in (\ref{current_zeta}) is defined by using $\bar{g}_{\mu\nu}$.
For the current case of the 4D extremal Kerr black hole,
we have the charges $\{Q_{\zeta_n}\}$ associated with $\{\zeta_n\}$,
and also $Q_{\partial_{\tau}}$, which is associated with
$\partial_{\tau}$.
Since $Q_{\partial_{\tau}}$ measures the deviation from the extremality,
it is fixed to zero in this case.
That is, we put a Dirac constraint $Q_{\partial_{\tau}}=0$ by hand.

Under this constraint $Q_{\partial_{\tau}}=0$,
the Dirac brackets of the charges are,
by considering the transformation property of the charge
$Q_{\zeta}$ under the diffeomorphism generated by $\xi$,
given as
\begin{align}
\{Q_{\zeta},Q_{\xi}\}_{\mathit{Dirac}}
&=Q_{[\zeta,\xi]_{\mathit{Lie}}}
+\frac{1}{8\pi G_d}\int_{\partial \Sigma}
k_{\zeta}[\mathcal{L}_{\xi}\bar{g},\bar{g}].
\label{dirac_algebra}
\end{align}

For $\{Q_{\zeta_n}\}$,
using \eqref{dirac_algebra},
the Dirac brackets are
\begin{align}
\{Q_{\zeta_{m}},Q_{\zeta_n}\}_{\mathit{Dirac}}
&=Q_{[\zeta_{m},\zeta_{n}]_{\mathit{Lie}}}
+\frac{1}{8\pi G_4}\int_{\partial \Sigma}
k_{\zeta_{m}}[\mathcal{L}_{\zeta_{n}}\bar{g},\bar{g}] \nonumber\\
&=Q_{[\zeta_{m},\zeta_{n}]_{\mathit{Lie}}}-iJm(m^2+2)\delta_{m+n,0}.
\label{kerr_dirac_algebra}
\end{align}
By redefining the charges as $L_n=Q_{\zeta_n}+3J\delta_{n,0}/2$
and replacing the Dirac bracket $\{.,.\}_{\mathit{Dirac}}$
by a commutator $-i[.,.]$,
we see that $\{L_n\}$ satisfies a Virasoro algebra
\begin{align}
[L_m,L_n] =(m-n)L_{m+n}+\frac{c}{12}m(m^2-1)\delta_{m+n,0},
\label{virasoro_central}
\end{align}
with the central charge $c=12J$.

The temperature of the dual chiral CFT$_2$ is, on the other hand,
determined by
identifying quantum numbers in the near horizon geometry
with those in the original geometry \cite{Frolov:1989jh}.
For this purpose let us consider a free scalar field $\Phi$
propagating on the nonextremal Kerr black hole background.
It can be expanded in eigenmodes of the asymptotic energy
$\omega$ and angular momentum $m$ as
\begin{align}
\Phi = \displaystyle{\sum_{\omega,m,l}}
\Phi_{\omega,m,l}e^{-i\omega\hat{t}+im\hat{\phi}}f_l(r,\theta).
\end{align}
Here $f_l(r,\theta)$'s are spherical harmonics.
Using the coordinate transformation \eqref{very_near_kerr},
we see that
\begin{align}
& e^{-i\omega\hat{t}+im\hat{\phi}}\Phi = e^{-in_{\tau}t+in_{\varphi}\phi}\Phi, \\
& n_{\tau}=m, \qquad n_{\varphi}  = \frac{1}{\epsilon}(2M\omega-m).
\end{align}
By tracing out the states inside the horizon,
the vacuum state includes the Boltzmann factor,
and the temperatures are determined by
\begin{align}
& e^{-\frac{\omega -m\Omega_H}{T_H}}
= e^{-\frac{n_{\varphi}}{T^{\varphi}}-\frac{n_{\tau}}{T^{\tau}}}, \\
& T^{\varphi}=\frac{r_+-M}{2\pi(r_+-a)},
\qquad T^{\tau}=\frac{r_+-M}{2\pi\epsilon r_+}.
\end{align}
and taking the near horizon extremal limit
\eqref{kerr_near_horizon_extremal_limit},
we see that $T^{\tau}$ vanishes while $T^{\varphi}$ is
\begin{align}
T^{\varphi}=\frac{1}{2\pi}.
\label{kerr_temperature}
\end{align}

We can also calculate the temperature of the chiral CFT$_2$ by
starting with the 1st law of thermodynamics:
\begin{align}
dS_{\mathit{BH}} = \beta_H dM - \tilde{\beta} dJ,
\label{kerr_1st_law}
\end{align}
where $\beta_H=1/T_H$ is the inverse Hawking temperature and
$\tilde{\beta} = \beta_H\Omega_H $.
Here we notice that the entropy $S_{\mathit{BH}}(M,J)$
is a function of $M$ and $J$.
Using the nonextremality parameter $\delta$ \eqref{kerr_nonextremality},
the entropy is written as a function
$S_{\mathit{BH}}=\tilde{S}_{\mathit{BH}}(\delta,J)$ of
$\delta$ and $J$. Then the 1st law \eqref{kerr_1st_law} is rewritten as
\begin{align}
dS_{\mathit{BH}} = \beta_H d\delta + \beta dJ,
\label{1st_law_delta}
\end{align}
where
\begin{align}
\beta
= \Bigl(\frac{\partial \tilde{S}_{\mathit{BH}}}{\partial J}\Bigr)_{\delta:\text{fixed}}
=\frac{\beta_H}{2}\sqrt{\frac{G_4}{J}}-\tilde{\beta}
=\beta_H\Bigl(\frac{1}{2}\sqrt{\frac{G_4}{J}}-\Omega_H\Bigr).
\end{align}
When we impose the Dirac constraint $Q_{\partial_{\tau}}=0$, we only
consider deviations which preserve the extremality $\delta=0$.
Therefore the first term of \eqref{1st_law_delta} vanishes
and the inverse temperature $\beta$ of the dual chiral CFT$_2$
is calculated as
\begin{align}
\beta|_{\delta=0} =
\Bigl(\frac{\partial\tilde{S}_{\mathit{BH}}}{\partial J}\Bigr)_{\delta=0:\text{fixed}}
=2\pi.
\end{align}
That is, the temperature is $T^{\varphi}=\frac{1}{2\pi}$.
Here notice that the angular momenta are quantized by 1, not 1/2,
for scalar fields.
This result reproduces the temperature \eqref{kerr_temperature}
calculated above.

For more generic cases in which there are some quantized charges
$\{Q_i\}$ with the potentials $\{\Phi_i\}$
in addition to the mass $M$, we can calculate the temperature
in the same way. The 1st law is
\begin{align}
dS_{\mathit{BH}} &= \beta_H dM - \sum_{i}\tilde{\beta}_idQ_i, \\
\tilde{\beta_i} &= \beta_H\Phi_i.
\end{align}
Let us assume that the extremality condition is
written as $M=f(Q_i)$ where $f(Q_i)$ is a function of $\{Q_i\}$.
In this case, by defining $\delta=M-f(Q_i)$ and writing
the entropy as $S_{\mathit{BH}}=\tilde{S}_{\mathit{BH}}(\delta, Q_i)$,
the 1st law is written as
\begin{align}
dS_{\mathit{BH}} &= \beta_{H}d\delta + \sum_i\beta_idQ_i, \\
\beta_i &= \Bigl(\frac{\partial \tilde{S}_{\mathit{BH}}}{\partial Q_i}
\Bigr)_{\delta,{\,}Q_j{\,}(j\neq i):\text{fixed}}.
\end{align}
Then in the same way as above, the inverse temperature is
calculated as
\begin{align}
\beta_i|_{\delta=0}=
\Bigl(\frac{\partial \tilde{S}_{\mathit{BH}}}{\partial Q_i}
\Bigr)_{\delta=0,{\ }Q_j{\,}(j\neq i):\text{fixed}}.
\end{align}
We can easily show that this $\beta_i$ coincides with the one
calculated according to \cite{Guica:2008mu},
assuming that $\partial_t$ measures the nonextremality
where $t$ is the time of the near horizon AdS$_2$.
This method of calculating the temperatures has some merits.
We need only the Bekenstein-Hawking entropy formula for
the extremal black holes, and it is manifest that
$\beta_i$'s depend only on the very near horizon geometry.

As we have calculated the central charge and the temperature,
the entropy of the 4D Kerr black hole is microscopically
calculated via the thermal Cardy formula as
\begin{align}
S_{\mathit{micro}}=\frac{\pi^2}{3}cT^{\varphi} = 2\pi J,
\end{align}
which agrees with the Bekenstein-Hawking entropy \eqref{kerr_sbh}
with the extremal condition \eqref{kerr_extremal_condition}.%
\footnote{
For general cases, the Bekenstein-Hawking entropy agrees with the
entropy calculated from the Cardy formula
if $c=  \frac{3}{2\pi^2}\frac{\partial({S_{\mathit{extr}}}^2)}{\partial Q_i}$,
where $S_{\mathit{extr}}(Q_i) \equiv \tilde{S}(\delta=0,Q_i)$.
}

Here we stress that
this dual theory is not like the CFT's in the usual AdS/CFT correspondence,
because $\varphi$ is a spacelike coordinate and
the Virasoro algebra does not contain the time translation generator.
Moreover, the isometry of the near horizon geometry does not
contain the $SL(2,\R)$ of the Virasoro algebra.

For the case of the 4D extremal Kerr black hole, there is only
one parameter $J$ and embedding in string theory is nontrivial.
These facts make it difficult for us to understand
the Kerr/CFT correspondence well. In \cite{Azeyanagi:2008kb}, this
holography is applied to the near horizon geometry
of rotating Kaluza-Klein black holes.
They are realized as a rotating D0-D6 bound state in string theory.
In this case, there are two boundary conditions for each of which
one Virasoro symmetry is included in the ASG.
An essential point is that, for
one geometry, we can identify different and consistent
ASG's by taking different boundary conditions appropriately.
In other words, there are two ways to realize
a holographic dual chiral CFT$_2$ for one geometry.
This seems to be different from U-duality {\cite{Horowitz:1996ay}}
in that the geometry itself is unique in this case.

In this paper, in order to understand the Kerr/CFT correspondence well,
we will apply this correspondence to very well-known black holes
embedded in string theory.

\section{Near Horizon Holography in the AdS$_3$ Throat}
\label{near_horizon}

In this section
we take the ``near horizon limit'' \cite{Maldacena:1997re}
for the rotating D1-D5-P system.
This is the first decoupling limit, which is
familiar in the context of AdS/CFT correspondence.
In this limit, we find
an AdS$_3$ throat structure and
can apply the Brown-Henneaux's method 
to calculate the entropy microscopically.
In contrast to the Kerr/CFT-like holographies,
the dual CFT$_2$ for this region turns out to be nonchiral.

\subsection{The rotating D1-D5-P black holes}
\label{general_rotating}

The main object we will consider in this paper is
the rotating D1-D5-P black holes
\cite{Cvetic:1996xz,Cvetic:1998xh}
in type IIB supergravity or superstring.
They include the well-known BMPV
\cite{Breckenridge:1996is,Breckenridge:1996sn}
and the nonrotating D1-D5-P
\cite{Strominger:1996sh,Callan:1996dv,Maldacena:1996ky}
black holes as special cases.

\subsubsection{The supergravity solution}

A rotating D1-D5-P black hole is a solution
of 10D type IIB supergravity
on the compactified background $\R^{1,4}\times S^1\times T^4$,
which is regarded as a 5D spacetime macroscopically.
The 10D metric in the string frame
and the dilaton field are written as follows:
\begin{align}
ds^2_{str} &=
ds_6^2 + \sqrt{\frac{H_1}{H_5}}\,(dx_6^2+dx_7^2+dx_8^2+dx_9^2),\\
\label{6D_metric}
ds_6^2 &=
\frac{1}{\sqrt{H_1 H_5}}
\biggl[
-\Bigl(1-\frac{2mf_D}{\hat{r}^2}\Bigr)d\tilde{t}^2+d\tilde{y}^2
+H_1H_5f_D^{-1}
\frac{\hat{r}^4}{(\hat{r}^2+l_1^2)(\hat{r}^2+l_2^2)-2m\hat{r}^2}d\hat{r}^2
\nonumber \\
&\quad -\frac{4mf_D}{\hat{r}^2}\cosh\alpha_1\cosh\alpha_5
(l_2\cos^2\theta d\hat{\psi}+l_1\sin^2\theta d\hat{\phi})d\tilde{t}
\nonumber \\
&\quad -\frac{4mf_D}{\hat{r}^2}\sinh\alpha_1\sinh\alpha_5
(l_1\cos^2\theta d\hat{\psi}+l_2\sin^2\theta d\hat{\phi})d\tilde{y}
\nonumber \\
&\quad +\Bigl(
\bigl(1+\frac{l_2^2}{\hat{r}^2}\bigr)H_1H_5\hat{r}^2+(l_1^2-l_2^2)\cos^2\theta
\bigl(\frac{2mf_D}{\hat{r}^2}\bigr)^2\sinh^2\alpha_1\sinh^2\alpha_5
\Bigr)\cos^2\theta d\hat{\psi}^2 \nonumber \\
&\quad  +\Bigl(
\bigl(1+\frac{l_1^2}{\hat{r}^2}\bigr)H_1H_5\hat{r}^2+(l_2^2-l_1^2)\sin^2\theta
\bigl(\frac{2mf_D}{\hat{r}^2}\bigr)^2\sinh^2\alpha_1\sinh^2\alpha_5
\Bigr)\sin^2\theta d\hat{\phi}^2 \nonumber \\
&\quad +\frac{2mf_D}{\hat{r}^2}(l_2\cos^2\theta d\hat{\psi} 
+l_1 \sin^2\theta d\hat{\phi})^2 +H_1H_5\hat{r}^2f_D^{-1}d\theta^2
\biggr],\\
e^{-2\Phi} &= \frac{1}{g_s^2}\frac{H_5}{H_1},
\end{align}
where
\begin{align}
H_1 &= 1+\frac{2mf_D \sinh^2\alpha_1}{\hat{r}^2},
\quad H_5 = 1+\frac{2mf_D \sinh^2\alpha_5}{\hat{r}^2},\\
f_D^{-1} &= 1+\frac{l_1^2\cos^2\theta+l_2^2\sin^2\theta}{\hat{r}^2},
\end{align}
and
\begin{align}
d\tilde{t} &= \cosh \alpha_p d\hat{t}-\sinh \alpha_p d\hat{y}, \\
d\tilde{y} &= \cosh \alpha_p d\hat{y}-\sinh \alpha_p d\hat{t}.
\end{align}
Here $x_6,\dots, x_9$ are the coordinates of the $T^4$
with $x_a\simeq x_a+2\pi R_a$ ($a=6,\dots, 9$),
and $\hat{y}$ is also compactified as $\hat{y}\simeq \hat{y}+2\pi R$.
For later convenience we define $V=R_6R_7R_8R_9$.
After the dimensional reduction of the $T^4$,
we get a solution of the Kaluza-Klein (KK) compactified
6D $\mathcal{N}=4$ (maximally supersymmetric)
supergravity, whose metric in the 6D Einstein frame is identical with
\eqref{6D_metric}.

The six parameters $m$, $l_1$, $l_2$, $\alpha_1$, $\alpha_5$, $\alpha_p$ are 
related to the conserved charges as 
\begin{align}
\label{charges_start}
M &= m\sum_{i=1,5,p}\cosh2\alpha_i , \\
Q_1 &= m\sinh 2\alpha_1,
\quad Q_5 = m\sinh 2\alpha_5,
\quad Q_p = m\sinh 2\alpha_p,\\
J_L = \frac{1}{2}(J_{\phi}-J_{\psi})
&= \frac{\pi}{4G_5}
m(l_1-l_2)\Bigl(\prod_{i=1,5,p}\cosh\alpha_i
+\prod_{i=1,5,p}\sinh\alpha_i\Bigr), \\
J_R = \frac{1}{2}(J_{\phi}+J_{\psi})
&= \frac{\pi}{4G_5}
m(l_1+l_2)\Bigl(\displaystyle{\prod_{i=1,5,p}\cosh\alpha_i}
-\prod_{i=1,5,p}\sinh\alpha_i\Bigr).
\label{charges_end}
\end{align}
Here the 5D, 6D and 10D Newton constants $G_5$, $G_6$, $G_{10}$ are
related with one another as
\begin{align}
(2\pi)^5VR\,G_5 = (2\pi)^4V G_6 = G_{10}.
\end{align}
Actually,
this solution represents a black hole geometry
if and only if
\begin{align}
m \ge \frac{1}{2}(|l_1| + |l_2|)^2,
\label{black_hole_condition}
\end{align}
and otherwise it becomes a smooth soliton,
a conical defect or a naked singularity \cite{Jejjala:2005yu}.
We always assume \eqref{black_hole_condition} below in this paper.
Then the outer/inner horizons are located at
\begin{align}
\label{horizons}
\hat{r}_{\pm}^2 =
\frac{2m-(l_1^2+l_2^2)}{2}
\pm \frac{1}{2}\sqrt{[2m-(l_1^2+l_2^2)]^2-4l_1^2l_2^2},
\end{align}
and the Bekenstein-Hawking entropy is calculated as 
\begin{align}
\label{general_Sbh}
S_{BH}&= \frac{\pi^2 m}{2G_5}
\biggl[
\Bigl(\prod_{i=1,5,p}\cosh\alpha_i+
\prod_{i=1,5,p}\sinh\alpha_i\Bigr)
\sqrt{2m-(l_1-l_2)^2} \nonumber \\
&\qquad\;\;\,
+\Bigl( \prod_{i=1,5,p}\cosh\alpha_i-
\prod_{i=1,5,p}\sinh\alpha_i\Bigr)
\sqrt{2m-(l_1+l_2)^2}
\,\biggr].
\end{align}

In string theory, the charges are quantized and they can be 
written as 
\begin{align}
\label{N1_quantize}
Q_1&=c_1N_1,
\quad c_1=\frac{g_s\alpha'^3}{V},
\quad N_1=\frac{V m}{g_s\alpha'^3}\sinh2\alpha_1,\\
Q_5&=c_5N_5,
\quad c_5=g_s\alpha',
\;\;\quad N_5=\frac{m}{g_s\alpha'}\sinh2\alpha_5, \\
Q_p&=c_pN_p,
\quad c_p=\frac{g_s^2\alpha'^4}{VR^2},
\quad N_p=\frac{VR^2m}{g_s^2\alpha'^4}\sinh2\alpha_p,
\end{align}
with $N_1$, $N_5$, $N_p$ integers.
$J_\phi$ and $J_\psi$ are also quantized as usual,
using integers $N_\phi$, $N_\psi$:
\begin{align}
J_\phi &= J_L+J_R=\frac{1}{2}N_\phi, \\
J_\psi &= J_R-J_L=\frac{1}{2}N_\psi.
\label{Jpsi_quantize}
\end{align}
The 10D Newton constant $G_{10}$ is expressed as
\begin{align}
G_{10} &= 8\pi^6 g_s^2 \alpha'^4.
\end{align}

\subsubsection{D-brane configuration and the D1-D5 CFT}
\label{d1d5_cft}

In type IIB superstring,
this system corresponds to a bound state of D-branes.
In that picture,
$N_5$ D5-branes wrap around the $S^1\times T^4$,
$N_1$ D1-branes wrap around the $S^1$
and $N_p$ Kaluza-Klein momenta run along the $S^1$ direction.
That is the reason why this system is called D1-D5-P.
It is shown in Table \ref{d1d5p_table}.

\begin{table}[h]
\centering
\begin{tabular}{ccccc}
 & $t$ &$\R^{4}$ & $S^1$ & $T^4$ \\
\hline
D1 & $\bigcirc$ &   & $\bigcirc$ & \\
D5 & $\bigcirc$ &   & $\bigcirc$ &
     $\bigcirc$ $\bigcirc$ $\bigcirc$ $\bigcirc$ \\
P  & $\bigcirc$ &   & $\bigcirc$ &
\end{tabular}
\caption{The D1-D5-P bound state}
\label{d1d5p_table}
\end{table}

In this system,
the open strings on the branes decouple
from the bulk closed strings in the low energy limit
and are described by a 2D $U(N_1)\times U(N_5)$ $\mathcal{N}=(4,4)$
superconformal field theory,
which is called the D1-D5 CFT.
This is the representative example of the
AdS$_3$/CFT$_2$ correspondence.
The dominant part of the degrees of freedom
comes from the fields of the bifundamental representation
of the $U(N_1)\times U(N_5)$.
This yields the central charges
\begin{align}
c^L = c^R = 6N_1N_5.
\end{align}

The angular momenta $J_L$, $J_R$ correspond to the R-charges
of the R-symmetry $SO(4)_R\simeq SU(2)_L\times SU(2)_R$.
Thus in order that the black hole have angular momenta,
the fields of the adjoint representation of the $U(N_1)$ or $U(N_5)$ 
also have to be excited as well as the bifundamental fields.
Still in this case, the leading term of the statistical entropy
is given by counting the configurations of the bifundamental fields,
and in fact it reproduces the Bekenstein-Hawking entropy \eqref{general_Sbh}
\cite{Breckenridge:1996is,Breckenridge:1996sn}.

\subsubsection{The extremal and the BPS conditions}
\label{extremal_and_bps}

This solution represents an extremal black hole
when the two horizons \eqref{horizons} coincide.
Remembering \eqref{black_hole_condition},
this occurs if and only if
$m = \frac{1}{2}(|l_1| + |l_2|)^2$.
We also take $l_1,l_2\ge 0$ for simplicity%
\footnote{
This is always possible without changing the geometry,
by the redefinitions of the parameters and the coordinates,
$(l_1,\alpha_p,y,\phi)\to(-l_1,-\alpha_p,-y,-\phi)$
\;or\;
$(l_2,\alpha_p,y,\psi)\to(-l_2,-\alpha_p,-y,-\psi)$.
}
and then the extremal condition is written as
\begin{align}
m=\frac{(l_1+l_2)^2}{2},
\label{extremal_condition}
\end{align}
and at that time the horizon is located at
\begin{align}
\hat{r}_H = \sqrt{l_1l_2}.
\end{align}
Under this condition,
the Bekenstein-Hawking entropy \eqref{general_Sbh} reduces to
\begin{align}
\label{extremal_Sbh}
S_{BH}
&= 2\pi \sqrt{\Bigl(\frac{\pi}{4G_5}\Bigr)^2Q_1Q_5Q_p-J_L^2+J_R^2} \nonumber\\
&= 2\pi \sqrt{N_1N_5N_p+\frac{N_{\phi}N_{\psi}}{4}}.
\end{align}
Here we used \eqref{N1_quantize}-\eqref{Jpsi_quantize}.

Note that \eqref{extremal_condition} is the {\it extremal} condition,
and it does not necessarily mean
that the solution is supersymmetric or BPS saturated.
The BPS bound is written as
\begin{align}
M \ge Q_1+Q_5+Q_p.
\label{BPS_bound}
\end{align}
In fact, this BPS condition is easily derived from
\eqref{charges_start}-\eqref{charges_end}.
If we write the parameters as
\begin{align}
e^{\alpha_i} = \frac{\eta_i}{\sqrt{m}} \;\; (i=1,5,p),
\qquad
l_a = \sqrt{m}\,j_a \;\; (a=1,2),
\label{l_for_BPS}
\end{align}
then the BPS limit is given by $m\to 0$, $\alpha_i\to\infty$
while $\eta_i$, $j_a$ and $G_{10}$, $V$, $R$ are fixed.
In this limit, the BPS bound
\eqref{BPS_bound} is saturated and
the metric reduces to a rather simple form,
\begin{align}
\label{6DBMPV_metric}
ds_6^2 &=
\frac{1}{\sqrt{H_1 H_5}}
\Bigl[
-d\hat{t}^2+d\hat{y}^2 + H_p(d\hat{t}-d\hat{y})^2
+H_1H_5(d\hat{r}^2 + \hat{r}^2\,d\Omega_3^2) \nonumber\\
&\qquad\qquad\quad
- \frac{8G_5J_L}{\pi\hat{r}^2}
(\sin^2\theta\,d\hat{\phi}
- \cos^2\theta\,d\hat{\psi})(d\hat{t}-d\hat{y}) \Bigr],\\
H_1 &= 1+\frac{Q_1}{\hat{r}^2},
\qquad H_5 = 1+\frac{Q_5}{\hat{r}^2},
\qquad H_p = \frac{Q_p}{\hat{r}^2},
\end{align}
and we also get $J_R=0$.
This is nothing but the BMPV black hole
with three different charges.
Under \eqref{l_for_BPS},
the extremal condition \eqref{extremal_condition}
is expressed as
\begin{align}
\label{extremal_BPS_condition}
j_1 + j_2 = \sqrt{2}.
\end{align}
But at this time the solution does not include $j_1+j_2$,
therefore we can always satisfy \eqref{extremal_BPS_condition}.
This shows that the BPS black hole (BMPV) is just a special one
among more general extremal D1-D5-P black holes.%
\footnote{
The nonrotating D1-D5-P black hole was obtained if we further set $j_1=j_2$.
}
In fact, general supersymmetric black holes with regular horizons
are proved to be extremal in \cite{Kunduri:2008tk}.

\label{instabilities}
Apart from BMPV, the extremal rotating D1-D5-P black holes are
not BPS saturated.
The BPS states with the same charges and angular momenta as them
are known to be black rings
\cite{Elvang:2004rt,Bena:2004de,Elvang:2004ds,Gauntlett:2004qy}.
Then these black holes are not really stable
even though they are extremal,
and they could be expected to decay to the black rings
through a tunneling process after a very long time.
Moreover, they have an ergoregion outside the horizon.
It leads to so-called superradiance instability,
in which the black hole emits its mass together with its angular momenta
simultaneously.
(The D-brane picture of this instability was proposed in \cite{Dias:2007nj}.)

Considering holographic duals of such unstable background geometries
obviously includes some subtle problems.
But at least for the microstate counting, symmetries and related issues,
we can reasonably expect that it does not affect our discussions later.

\newpage
\subsection{The AdS$_3$ decoupling limit}

Here we take the near horizon decoupling limit \cite{Maldacena:1997re},
in which the degrees of freedom in the AdS$_3$ throat
will decouple from the asymptotic flat region.

If we define $\lambda=(Q_1Q_5)^{1/4}$,
then in terms of quantities in string theory,
\begin{align}
\lambda
= \Bigl(\frac{g_s^2\alpha'^4}{V}N_1N_5 \Bigr)^{1/4}
= \Bigl(\frac{g_s^2}{v}N_1N_5 \Bigr)^{1/4}l_s,
\end{align}
where we put $V=\alpha'^2 v$ and $l_s=\sqrt{\alpha'}$.
Now we take the decoupling limit as
\begin{align}
l_s\to 0 \quad \text{with}
\quad
\begin{array}{l}
\displaystyle
g_s,\quad v,\quad R,
\quad U=\frac{\hat{r}}{\lambda^2},
\quad c=\frac{m}{\lambda^4},
\quad \alpha_p, \\
\displaystyle
w_i = \lambda e^{\alpha_i}\;\;(i=1,5),
\quad
b_a=\frac{l_a}{\lambda^2}\;\;(a=1,2)
\end{array}
\quad \text{fixed}.
\label{ads3_decoupling_limit}
\end{align}
In this limit, the Newton constants also scale as
\begin{align}
G_{10} \sim l_s^8,
\qquad
G_6 \sim l_s^4,
\qquad
G_5 \sim \frac{l_s^4}{R},
\end{align}
with the angular momenta $J_L$, $J_R$ and
the quantized charges $N_1, N_5, N_p$ remaining finite.

Under these scalings, we obtain the near horizon geometry
\begin{align}
\label{near_horizon_geometry}
\frac{ds_6^2}{\lambda^2} &= \frac{U^2}{f_D}
\Bigl(-\bigl(1-\frac{2c f_D}{U^2}\bigr)d\tilde{t}^2+
d\tilde{y}^2\Bigr)
+\frac{U^2}{(U^2+b_1^2)(U^2+b_2^2)-2cU^2}dU^2 \nonumber \\
&\quad - 2(b_2\cos^2\theta d\hat{\psi} +b_1\sin^2\theta d\hat{\phi})d\tilde{t}
-2(b_1\cos^2\theta d\hat{\psi} 
+b_2\sin^2\theta d\hat{\phi})d\tilde{y} \nonumber \\
&\quad +  (d\theta^2+\sin^2\theta d\hat{\phi}^2+\cos^2\theta d\hat{\psi}^2),
\end{align}
with 
\begin{align}
f_D^{-1} = 1+\frac{b_1^2\sin^2\theta+b_2^2\cos^2\theta}{U^2}.
\end{align}
By introducing 
\begin{align}
d\tilde{\psi} &= d\hat{\psi} -(b_2 d\tilde{t}+ b_1 d\tilde{y}), \\
d\tilde{\phi} &= d\hat{\phi} -(b_1 d\tilde{t}+b_2 d\tilde{y}),
\end{align}
it can be rewritten as 
\begin{align}
\frac{ds_6^2}{\lambda^2} 
&= -\frac{(U^2+b_1^2)(U^2+b_2^2)-2cU^2}{U^2}d\tilde{t}^2+
U^2\Bigl(d\tilde{y}-\frac{b_1b_2}{U^2}d\tilde{t}\Bigr)^2
\nonumber \\
&\quad +\frac{U^2}{(U^2+b_1^2)(U^2+b_2^2)-2cU^2}dU^2
+(d\theta^2+\sin^2\theta d\tilde{\phi}^2
+\cos^2\theta d\tilde{\psi}^2).
\end{align}
This metric is further rewritten in the standard BTZ form as 
\begin{align}
\frac{ds_6^2}{\lambda^2}
&= -N^2dt_{\mathit{BTZ}}^2 +N^{-2}dr_{\mathit{BTZ}}^2 
+r_{\mathit{BTZ}}^2(dy_{\mathit{BTZ}}-N_{y}dt_{\mathit{BTZ}})^2 \nonumber \\
&\quad + (d\theta^2+\sin^2\theta d\tilde{\phi}^2+\cos^2\theta d\tilde{\psi}^2),
\end{align}
where 
\begin{align}
N^2 &= r_{\mathit{BTZ}}^2-M_{\mathit{BTZ}}+\frac{16G_3^2J_{\mathit{BTZ}}^2}{r_{\mathit{BTZ}}^2}, \\
N_{y} &= \frac{4G_3J_{\mathit{BTZ}}}{r_{\mathit{BTZ}}^2},
\end{align}
with new coordinates 
\begin{align}
t_{\mathit{BTZ}} &= \frac{\hat{t}}{R},
\qquad y_{\mathit{BTZ}}= \frac{\hat{y}}{R}, \nonumber\\ 
r_{\mathit{BTZ}}^2 &= R^2(U^2+(2c-b_1^2-b_2^2)\sinh^2\alpha_p
+2b_1b_2\sinh\alpha_p\cosh\alpha_p),
\end{align}
where $y_{\mathit{BTZ}}$ is compactified
as $y_{\mathit{BTZ}}\sim y_{\mathit{BTZ}}+2\pi$.
The 3D Newton constant $G_3$ is
\begin{align}
G_3 &= \frac{G_6}{2\pi^2\lambda^3}.
\end{align}
The mass and the angular momentum of the BTZ black hole are
\begin{align}
M_{\mathit{BTZ}} &= R^2((2c-b_1^2-b_2^2)\cosh2\alpha_p+2b_1b_2\sinh2\alpha_p), \\
8G_3J_{\mathit{BTZ}}&= R^2((2c-b_1^2-b_2^2)\sinh2\alpha_p+2b_1b_2\cosh2\alpha_p),
\end{align}
and the horizon is given by $r_{\mathit{BTZ}} = r_{+}$, where
\begin{align}
r_{+}^2
= \frac{M_{\mathit{BTZ}}}{2}
+ \frac{1}{2}\sqrt{M_{\mathit{BTZ}}^2-(8G_3J_{\mathit{BTZ}})^2}.
\end{align}

It is convenient to adopt a further coordinate transformation
\begin{align}
\rho^2
= r_{\mathit{BTZ}}^2 - \frac{M_{\mathit{BTZ}}}{2}
+ \frac{1}{2}\sqrt{M_{\mathit{BTZ}}^2-(8G_3J_{\mathit{BTZ}})^2},
\end{align}
which leads to
\begin{align}
\label{near_horizon_geometry_final}
\frac{ds_6^2}{\lambda^2}
&= -N^2dt_{\mathit{BTZ}}^2 + \frac{\rho^2}{N^2\Xi}d\rho^2 
+\Xi\,(dy_{\mathit{BTZ}}-N_{y}dt_{\mathit{BTZ}})^2 \nonumber\\
&\quad
+ (d\theta^2+\sin^2\theta d\tilde{\phi}^2+\cos^2\theta d\tilde{\psi}^2),\\
N^2 &= \frac{\rho^2(\rho^2-\rho_+^2)}{\Xi},\\
N_{y} &= \frac{4G_3J_{\mathit{BTZ}}}{\Xi},
\end{align}
where
\begin{align}
\Xi &= r_{\mathit{BTZ}}^2 = \rho^2 + \frac{M_{\mathit{BTZ}}}{2}
 - \frac{1}{2}\sqrt{M_{\mathit{BTZ}}^2-(8G_3J_{\mathit{BTZ}})^2},\\
\rho_+^2 &= \sqrt{M_{\mathit{BTZ}}^2-(8G_3J_{\mathit{BTZ}})^2}.
\end{align}
Here the horizon is located at $\rho=\rho_+$.

\subsection{Asymptotic symmetry group and central charges}

For the near horizon geometry \eqref{near_horizon_geometry_final},
we can carry out the Brown-Henneaux's method.
In fact it was already done in \cite{Strominger:1997eq},
for the pure BTZ case without the $S^3$ fiber in this case.

In a similar manner as that,
the generators are expressed as
\begin{align}
\zeta &=
\Bigl[\Bigl(\frac{1}{2} + \frac{\lambda^2}{4\rho^2}\partial_R^2\Bigr)\gamma^{(R)}
+ \Bigl(\frac{1}{2} + \frac{\lambda^2}{4\rho^2}\partial_L^2\Bigr)\gamma^{(L)}
+ \mathcal{O}\Bigl(\frac{1}{\rho^4}\Bigr)
\Bigr]\partial_{t_{\mathit{BTZ}}} \nonumber\\
&\quad +
\Bigl[-\frac{\rho}{2\lambda}\partial_R\gamma^{(R)}
- \frac{\rho}{2\lambda}\partial_L\gamma^{(L)}
+ \mathcal{O}\Bigl(\frac{1}{\rho}\Bigr)
\Bigr]\partial_{\rho} \nonumber\\
&\quad +
\Bigl[ \Bigl(\frac{1}{2}-\frac{\lambda^2}{4\rho^2}\partial_R^2\Bigr)\gamma^{(R)}
- \Bigl(\frac{1}{2}-\frac{\lambda^2}{4\rho^2}\partial_L^2\Bigr)\gamma^{(L)}
+ \mathcal{O}\Bigl(\frac{1}{\rho^4}\Bigr)
\Bigr]\partial_{y_{\mathit{BTZ}}},
\end{align}
where
\begin{align}
x^R = t_{\mathit{BTZ}} + y_{\mathit{BTZ}},
\qquad
x^L = t_{\mathit{BTZ}} - y_{\mathit{BTZ}},
\end{align}
and $\gamma^{(R)}$ and $\gamma^{(L)}$ are arbitrary functions of
$x^R$ and $x^L$ respectively.

Now we define $\gamma^{(R)}_n=e^{inx^R}$,  $\gamma^{(L)}_n=e^{inx^L}$, and
\begin{align}
\label{near_horizon_generators_R}
\zeta^{(R)}_n &=
\Bigl[\Bigl(\frac{1}{2} + \frac{\lambda^2}{4\rho^2}\partial_R^2\Bigr)
\gamma^{(R)}_n\Bigr] \partial_{t_{\mathit{BTZ}}}
- \Bigl(\frac{\rho}{2\lambda}\partial_R\gamma^{(R)}_n\Bigr) \partial_\rho
+ \Bigl[\Bigl(\frac{1}{2}-\frac{\lambda^2}{4\rho^2}\partial_R^2\Bigr)
\gamma^{(R)}_n\Bigr] \partial_{y_{\mathit{BTZ}}},\\
\label{near_horizon_generators_L}
\zeta^{(L)}_n &=
\Bigl[\Bigl(\frac{1}{2} + \frac{\lambda^2}{4\rho^2}\partial_L^2\Bigr)
\gamma^{(L)}_n\Bigr] \partial_{t_{\mathit{BTZ}}}
- \Bigl(\frac{\rho}{2\lambda}\partial_L\gamma^{(L)}_n\Bigr)\partial_\rho
- \Bigl[\Bigl(\frac{1}{2}-\frac{\lambda^2}{4\rho^2}\partial_L^2\Bigr)
\gamma^{(L)}_n\Bigr] \partial_{y_{\mathit{BTZ}}},
\end{align}
then we can show easily that they satisfy the Virasoro algebras
\begin{align}
i[\zeta^{(R)}_m,\zeta^{(R)}_n]_{\mathit{Lie}} &= (m-n)\zeta^{(R)}_{m+n},\\
i[\zeta^{(L)}_m,\zeta^{(L)}_n]_{\mathit{Lie}} &= (m-n)\zeta^{(L)}_{m+n},\\
i[\zeta^{(R)}_m,\zeta^{(L)}_n]_{\mathit{Lie}} &=
\mathcal{O}\Bigl(\frac{1}{\rho^4}\Bigr)\partial_{t_{\mathit{BTZ}}}
+ \mathcal{O}\Bigl(\frac{1}{\rho^4}\Bigr)\partial_{y_{\mathit{BTZ}}}.
\end{align}

The central extensions are computed as
\begin{align}
\frac{1}{8\pi G_6}\int_{\partial \Sigma}
k_{\zeta^{(R)}_{m}}[\mathcal{L}_{\zeta^{(R)}_{n}}\bar{g},\bar{g}]
&=
\frac{1}{8\pi G_6}\int_{\partial \Sigma}
k_{\zeta^{(L)}_{m}}[\mathcal{L}_{\zeta^{(L)}_{n}}\bar{g},\bar{g}] \nonumber\\
&=
-i\biggl(\frac{\pi^2\lambda^4}{4G_6}m^3
+\frac{\pi^2l_1l_2R^2(\cosh 2\alpha_p + \sinh 2\alpha_p)}{G_6}m\biggr)
\delta_{m+n,0}.
\label{near_horizon_central_extensions}
\end{align}
In this system the gauge and the scalar fields exist
other than the metric,
but their contribution to the central charges would vanish
\cite{Murata:2009}.
Then \eqref{near_horizon_central_extensions}
gives the central charges as
\begin{align}
c^R = c^L
&= \frac{3\pi^2\lambda^4}{G_6} \nonumber\\
&= 6N_1N_5.
\label{btz_central_charges}
\end{align}
Therefore in exactly the same manner as \cite{Strominger:1997eq},
in the whole parameter space of the charges and the angular momenta,
the dual CFT$_2$ is nonchiral and the central charges are the same
as the brane effective theory.
Therefore it seems to be natural that we would identify these
Virasoro symmetries to those of the CFT$_2$ on the D-branes.

\section{Very Near Horizon Limit and Kerr/CFT}
\label{very_near_horizon}

In the previous section,
we took the decoupling limit for the D1-D5-P system.
This is the ordinary near horizon limit in context of
the AdS/CFT correspondence,
and we were left with the AdS$_3$ throat structure.

In turn, at the bottom of this AdS$_3$ throat,
we find the BTZ black hole.
Therefore if it is extremal,
we can again go into the near horizon decoupling region,
which has an AdS$_2$ structure.
Following \cite{Strominger:1998yg},
we call it {\it very near horizon limit}.
In this limit we will naturally find the Kerr/CFT-like structure.

\subsection{Extremal limit and very near horizon geometry}

In order that the very near horizon limit can be taken consistently,
the BTZ black hole has to be extremal.
This in turn demands that the original D1-D5-P black hole should be extremal.

The extremal condition was given in
\eqref{extremal_condition},
and under \eqref{ads3_decoupling_limit}
it is described as
\begin{align}
c =\frac{(b_1+b_2)^2}{2}.
\end{align}
In this case $M_{\mathit{BTZ}}=8G_3J_{\mathit{BTZ}}$, $\rho_+=0$ and
the near horizon geometry \eqref{near_horizon_geometry} becomes
\begin{align}
\frac{ds_6^2}{\lambda^2}
&=
-\frac{\rho^4}{\rho^2+r_+^2}dt_{\mathit{BTZ}}^2
+ \frac{d\rho^2}{\rho^2}
+ (\rho^2 + r_{+}^2)(dy_{\mathit{BTZ}}
   -\frac{r_+^2}{\rho^2+r_+^2}dt_{\mathit{BTZ}})^2 \nonumber\\
&\quad +(d\theta^2+\sin^2\theta d\tilde{\phi}^2
+\cos^2\theta d\tilde{\psi}^2),
\label{btz_form_metric}
\end{align}
where
\begin{align}
r_+^2 = \frac{M_{\mathit{BTZ}}}{2} = R^2b_1b_2(\sinh2\alpha_p+\cosh2\alpha_p).
\end{align}

Now let us next take the very near horizon limit.
By defining
\begin{align}
\rho^2 &= 2\epsilon\, r, \quad t_{\mathit{BTZ}}=\frac{t}{\epsilon},
\quad y_{\mathit{BTZ}} = y+\frac{t}{\epsilon}, \nonumber\\
\hat{\phi} &= \phi
+\frac{t}{\epsilon}R(b_1+b_2)(\cosh\alpha_p-\sinh\alpha_p), \nonumber\\
\hat{\psi} &= \psi
+\frac{t}{\epsilon}R(b_1+b_2)(\cosh\alpha_p-\sinh\alpha_p),
\label{very_near_horizon_limit}
\end{align}
and taking $\epsilon \to 0$, we obtain the very near horizon 
geometry%
\footnote{
For the case of the nonrotating D1-D5-P black hole,
this is the same as the one obtained in \cite{Strominger:1998yg}.
}
\begin{align}
\frac{ds_6^2}{\lambda^2}
&=
- 4r^2dt^2 +\frac{dr^2}{4r^2}
+ r_+^2\bigl(dy+\frac{2r}{r_+}dt\bigr)^2
\nonumber\\
&\quad\; +d\theta^2
+ \sin^2\theta \bigl[d\phi+R(b_1\sinh\alpha_p-b_2\cosh\alpha_p)\,dy\bigr]^2
\nonumber\\*
&\qquad\quad\;\;\;
+ \cos^2\theta \bigl[d\psi+R(b_2\sinh\alpha_p-b_1\cosh\alpha_p)\,dy\bigr]^2
\nonumber\displaybreak[2]\\
&=
- 4r^2dt^2 +\frac{dr^2}{4r^2}
+ r_+^2\bigl(dy+\frac{2r}{r_+}dt\bigr)^2
\nonumber\\
&\quad\;
+ d\theta^2
+ \sin^2\theta \Bigl(d\phi - \frac{2G_6}{\pi^2\lambda^4}J_{\psi}\,dy\Bigr)^2
+ \cos^2\theta \Bigl(d\psi - \frac{2G_6}{\pi^2\lambda^4}J_{\phi}\,dy\Bigr)^2.
\label{very_near_metric_rotating}
\end{align}

Here note that, we could also take the very near horizon limit
directly for the black hole geometry \eqref{6D_metric}
with the extremal condition \eqref{extremal_condition},
by defining
\begin{align}
\hat{r}^2 &= l_1l_2 + \epsilon\chi,
\quad \hat{t}=\frac{t}{\epsilon}\,R,
\nonumber\\
\hat{y} &= \Bigl(y+
\frac{t}{\epsilon}\,
\frac{e^{2\alpha_1 + 2\alpha_5 + 2\alpha_p}
      -e^{2\alpha_1}-e^{2\alpha_5}+e^{2\alpha_p}}
     {e^{2\alpha_1 + 2\alpha_5 + 2\alpha_p}
      +e^{2\alpha_1}+e^{2\alpha_5}+e^{2\alpha_p}}
\Bigr)R,
\nonumber\\
\hat{\phi} &= \phi +
\frac{t}{\epsilon}\,
\frac{4R}{l_1+l_2}\,
\frac{e^{\alpha_1 + \alpha_5 + \alpha_p}}
     {e^{2\alpha_1 + 2\alpha_5 + 2\alpha_p}
      +e^{2\alpha_1}+e^{2\alpha_5}+e^{2\alpha_p}},
\nonumber\\
\hat{\psi} &= \psi +
\frac{t}{\epsilon}\,
\frac{4R}{l_1+l_2}\,
\frac{e^{\alpha_1 + \alpha_5 + \alpha_p}}
     {e^{2\alpha_1 + 2\alpha_5 + 2\alpha_p}
      +e^{2\alpha_1}+e^{2\alpha_5}+e^{2\alpha_p}},
\label{direct_very_near_coordinates}
\end{align}
and taking $\epsilon\to 0$.
After that,
taking the AdS$_3$ decoupling limit \eqref{ads3_decoupling_limit}
reproduces the same result as \eqref{very_near_metric_rotating},
under the further coordinate transformation
\begin{align}
\chi = 2r\frac{\lambda^4 r_+}{R^2}.
\end{align}
In particular, for the BPS (BMPV) geometry \eqref{6DBMPV_metric},
just taking the very near horizon limit
yields \eqref{very_near_metric_rotating},
without assuming \eqref{ads3_decoupling_limit}.
In this case the coordinate transformation is written as
\begin{align}
\hat{r}^2 &= 2\epsilon r\frac{\lambda^4 r_+}{R^2},
\quad \hat{t}=\frac{t}{\epsilon}\,R,
\quad \hat{y} = \Bigl(y+\frac{t}{\epsilon}\Bigr)R,
\quad \hat{\phi} = \phi,
\quad \hat{\psi} = \psi,
\end{align}
where
\begin{align}
\lambda &= (Q_1Q_5)^{1/4}
= \frac{j_1+j_2}{2}\sqrt{\eta_1\eta_5}, \\
r_+ &= \frac{\eta_p\sqrt{j_1j_2}}{\lambda^2}\,R.
\end{align}
Here remember that the angular momenta are expressed as
\begin{align}
J_{\phi} = -J_{\psi} = J_L
= \frac{\pi}{16G_5}(j_1-j_2)\,\eta_1\eta_5\eta_p.
\end{align}

\subsection{Asymptotic symmetry groups and central charges}

For the very near horizon geometry
\eqref{very_near_metric_rotating},
we can take three different asymptotic symmetry groups,
which correspond to three different boundary conditions, respectively.
\eqref{very_near_metric_rotating} has
three independent $U(1)$ isometries along $y$, $\phi$, and $\psi$,
respectively,
and in each of the ASG's only one of the $U(1)$'s
is enhanced to a Virasoro symmetry.
This is exactly the same situation as was observed
in \cite{Azeyanagi:2008kb}.

In the case where $U(1)_y$ is enhanced,
the generators of the ASG are given as%
\footnote{
$\partial_t$ is excluded from the ASG,
since we put the Dirac constraint $Q_{\partial_t}=0$ at the same time.
}
\begin{align}
\zeta^{(y)}_{\gamma} &= \gamma(y)\partial_y - r\gamma'(y)\partial_y,\\
\zeta^{(\phi)}&= -\partial_\phi.\\
\zeta^{(\psi)}&= -\partial_\psi.
\end{align}
Then defining
\begin{align}
\gamma_n &= -e^{-iny}, \\
\zeta^{(y)}_n &= \gamma_n\partial_y - r\gamma_n'\partial_r,
\end{align}
the generators $\{\zeta^{(y)}_n\}$ satisfy the Virasoro algebra
\begin{align}
[\zeta^{(y)}_m, \zeta^{(y)}_n]_{\mathit{Lie}} = -i(m-n)\zeta^{(y)}_{m+n}.
\end{align}
The rest of the cases are similar.

The central extension terms of the corresponding Dirac brackets are
\begin{align}
\label{central_extension_y}
\frac{1}{8\pi G_6}\int_{\partial \Sigma}
k_{\zeta^{(y)}_{m}}[\mathcal{L}_{\zeta^{(y)}_{n}}\bar{g},\bar{g}]
&=
-i\frac{\pi^2\lambda^4}{G_6}
\Bigl(\frac{m^3}{4} + r_+^2m\Bigr)\delta_{m+n,0},\\
\frac{1}{8\pi G_6}\int_{\partial \Sigma}
k_{\zeta^{(\phi)}_{m}}[\mathcal{L}_{\zeta^{(\phi)}_n}\bar{g},\bar{g}]
&=
-i \frac{\pi^2R\lambda^4}{4G_6}
\bigl(b_2\cosh\alpha_p - b_1\sinh\alpha_p\bigr)
m^3\delta_{m+n,0},\\
\frac{1}{8\pi G_6}\int_{\partial \Sigma}
k_{\zeta^{(\psi)}_m}[\mathcal{L}_{\zeta^{(\psi)}_n}\bar{g},\bar{g}]
&=
-i\frac{\pi^2R\lambda^4}{4G_6}
\bigl(b_1\cosh\alpha_p - b_2\sinh\alpha_p\bigr)
m^3\delta_{m+n,0},
\end{align}
which give the central charges as
\begin{align}
\label{very_near_center_y}
c^y &=\frac{3\pi^2\lambda^4}{G_6}
= \frac{3\pi^2Q_1Q_5}{G_6}
= 6N_1N_5,\\
\label{very_near_center_phi}
c^\phi &= \frac{3\pi^2R\lambda^4}{G_6}
(b_2\cosh\alpha_p - b_1\sinh\alpha_p)
= 6J_\psi = 12N_\psi, \\
\label{very_near_center_psi}
c^\psi &= \frac{3\pi^2R\lambda^4}{G_6}
(b_1\cosh\alpha_p - b_2\sinh\alpha_p)
= 6J_\phi = 12N_\phi,
\end{align}
respectively.
The results here are for the cases of
$N_1N_5, N_{\phi}, N_{\psi} > 0$, and
when, for example, $N_\phi < 0$,  
we can obtain a positive value for $c^{\psi}$
by a redefinition of the coordinate $\phi$.

It is notable that the right hand side of
\eqref{central_extension_y} is exactly the same as
\eqref{near_horizon_central_extensions}
and that $c^y$ is again the same value
as the brane effective theory and
the near horizon ASG, \eqref{btz_central_charges}.
The difference is that it is {\it chiral},
with only one Virasoro symmetry,
while the CFT on the D-branes and \eqref{btz_central_charges}
have both the left- and the right-movers.
Later we will discuss the underlying structure
behind this phenomenon.

\subsection{Temperatures and the microscopic entropy}

As we saw in \S\ref{review_kerr/cft},
the temperatures for the dual chiral CFT's
can be calculated from the macroscopic entropy formula
\eqref{extremal_Sbh}
for the extremal states.
It is straightforward that
\begin{align}
\label{very_near_Ty}
T^y &= \left(\frac{\partial S_{BH}}{\partial N_p}\right)^{-1}
= \frac{1}{\pi N_1N_5}\sqrt{N_1N_5N_p+\frac{N_{\phi}N_{\psi}}{4}},\\
T^\phi &= \left(\frac{\partial S_{BH}}{\partial (N_{\phi}/2)}\right)^{-1}
= \frac{1}{2\pi N_{\psi}} \sqrt{N_1N_5N_p+\frac{N_{\phi}N_{\psi}}{4}}, \\
\label{very_near_Tpsi}
T^\psi &= \left(\frac{\partial S_{BH}}{\partial (N_{\psi}/2)}\right)^{-1}
= \frac{1}{2\pi N_{\phi}} \sqrt{N_1N_5N_p+\frac{N_{\phi}N_{\psi}}{4}}.
\end{align}
Here again notice that the angular momenta are quantized by 1, not 1/2,
for scalar fields.

Substituting \eqref{very_near_center_y}-\eqref{very_near_center_psi}
and \eqref{very_near_Ty}-\eqref{very_near_Tpsi}
into the thermal Cardy formula,
we get
\begin{align}
S_{micro}
&= \frac{\pi^2}{3}c^yT^y
= \frac{\pi^2}{3}c^{\phi}T^{\phi}
= \frac{\pi^2}{3}c^{\psi}T^{\psi} \nonumber\\
&= 2\pi\sqrt{N_1N_5N_p+\frac{N_{\phi}N_{\psi}}{4}}.
\end{align}
Each of these agrees with the Bekenstein-Hawking entropy
\eqref{extremal_Sbh}.

\subsection{Virasoro symmetries and the CFT on the D-branes}

Now let us consider the relationships
of the very near horizon Virasoro symmetries
to the near horizon Virasoro symmetries
and those of the CFT on the D-branes.

In fact it is rather simple.
On the coordinate transformation \eqref{very_near_horizon_limit},
the Virasoro generators
\eqref{near_horizon_generators_R} and \eqref{near_horizon_generators_L}
become
\begin{align}
\zeta^{(R)}_n&\to
-\frac{n^2}{4r}e^{in(y+2t/\epsilon)}\partial_t
-irn e^{in(y+2t/\epsilon)}\partial_r
-\frac{n^2}{2r\epsilon} e^{in(y+2t/\epsilon)}\partial_y,
\\
\zeta^{(L)}_n&\to
-\frac{n^2}{4r}e^{-iny}\partial_t
-irn e^{-iny}\partial_r
-e^{-iny}\partial_y,
\label{zetaL_to_zetay}
\end{align}
under the very near horizon limit $\epsilon\to 0$.
Therefore
$\zeta^{(L)}_n$ turns to $\zeta^{(y)}_n$,
while $\zeta^{(R)}_0$ vanishes and
$\zeta^{(R)}_n (n\ne 0)$ diverges or vibrates infinitely fast.%
\footnote{
In fact there is a subtle problem about $\zeta^{(L)}_n$.
The first term of the right hand side of \eqref{zetaL_to_zetay}
is not an asymptotically trivial transformation.
However, we can show that
this term (with an arbitrary numerical factor)
does not affect the value of the the central charge $c^y=6N_0N_6$.
In addition,
under the Dirac constraint $Q_{\partial_t}=0$,
this term does not contribute to the transformation at all.
Therefore we can identify the right hand side of
\eqref{zetaL_to_zetay} with $\zeta^{(y)}_n$
for the current case.
}

It can be interpreted as follows.
The very near horizon limit corresponds to
a very low energy limit on the dual CFT$_2$,
in which we have only the {\it ground states} for fixed charges.%
\footnote{
As we explained in \S\ref{instabilities},
these ``ground states'' are not stable in truth for non-BPS extremal cases,
but it does not affect our discussions here and later.
}
The right-movers describe infinitely high energy excitations in this limit,
or in other words,
the mass gap is infinitely larger than the energy scale we focus on.
Therefore they have to decouple,
and finally we are left with a chiral Virasoro symmetry.

We can also regard it as the IR fixed point of the theory,
while the UV fixed point is the ordinary nonchiral D1-D5 CFT.
For the left-mover, we see from
\eqref{btz_central_charges} and \eqref{very_near_center_y}
that the central charges do not flow from the UV to the IR region.%
\footnote{
A similar comparison is carried out in \cite{Hotta:2008xt}
for a class of AdS black holes in a 3D Einstein-scalar theory.
They apply Brown-Henneaux's method at infinity (UV) and
at near horizon (IR), and show that
$c_{\mathit{IR}}<c_{\mathit{UV}}$ for that case.
}
Note that, at this IR fixed point,
states with different levels in a representation of this Virasoro algebra
correspond to different very near horizon backgrounds.
They are all extremal geometries with the same $N_1$ and $N_5$,
but have different $P$'s, as we will explain in the next section.

\section{The Kerr/CFT Correspondence as a Decoupling Limit}
\label{decoupling_limit}

For extremal black holes with a finite horizon area,
there is an AdS$_2$ structure with $U(1)^n$ fiber, which 
may imply that there exists a decoupling limit keeping the degrees of freedom
living deep in the AdS$_2$.
This limit is expected to be
the low energy limit to the lowest energy states,
namely the ground states with the fixed asymptotic charges.
Thus, the Kerr/CFT correspondence will be understood 
as the decoupling limit.

As we mentioned above,
in the D1-D5-P ($y$-direction) case,
the chiral CFT$_2$ does not live
on the very near horizon geometry with a fixed $P$.
We need more states than those on the one very near horizon geometry.
It might sound as if this statement contradicts with
the existence of the Virasoro symmetry in the ASG.
But notice that we now have $\zeta^{(y)}_0=-\partial_y$ in the ASG,
as a nontrivial transformation which is allowed by the boundary condition.
The corresponding charge to it is the KK momentum $P$,
and this means that we have many macroscopic geometries
with different $P$'s under the boundary condition.
For all of the CFT states corresponding to these geometries,
excitations of the right-movers are completely suppressed.
This should lead to the Dirac constraint $Q_{\partial_t}=0$
on the gravity side,
possibly by some quantum mechanism,
restricting the geometries to be extremal.
Therefore we find that the representation of the left-hand Virasoro algebra,
which is the set of the states of the dual chiral CFT$_2$,
corresponds to microstate geometries
which are all extremal and have different $P$'s.

On the other hand,
the central charge for the Virasoro algebra in the rotating direction
is roughly proportional to the angular momentum $J$ of the 
black hole.
Similar to the $y$-direction case,
the chiral CFT$_2$ is not expected to live in
the very near horizon geometry.
In addition to it, in this case
$J$ seems to parametrize the states rather than 
the boundary CFT in the AdS$_3$ throat,
while the central charge for the $y$-direction,
$c^y=6 N_1 N_5$, parametrizes the CFT$_2$.
Thus, it is natural to ask what is the origin of the chiral CFT$_2$
for the rotational direction.

As shown in
\cite{Kunduri:2007vf,Astefanesei:2007bf,Kunduri:2008rs,Kunduri:2008tk},
very near horizon geometries of
extremal black holes are highly constrained and then
we expect that 
there will be a kind of universality for such geometries.
An example of such universality,
shown in \cite{Kunduri:2008rs},
is the near horizon extremal geometries of
the 5D slow-rotating Kaluza-Klein black hole
\cite{Gibbons:1985ac,Rasheed:1995zv,Matos:1996km,Larsen:1999pp} and 
the 5D Myers-Perry black hole \cite{Myers:1986un}
on an orbifold $\R^4/\Z_{N_6}$,
which was shown in \cite{Kunduri:2008rs}.
Indeed, the 5D Myers-Perry black hole can be obtained 
from the KK black hole by taking a scaling limit 
\cite{Itzhaki:1998ka, Emparan:2007en},
thus the near horizon geometries should be the same.
Therefore,
the Kerr/CFT correspondence for the KK compactified direction 
of the KK black hole \cite{Azeyanagi:2008kb},
which is analogues to 
the $y$-direction in the D1-D5-P case,
is equivalent to the Kerr/CFT correspondence in the rotational 
direction of the 5D Myers-Perry black hole \cite{Lu:2008jk}.
The central charge for the compactified direction 
is $c^y=3 N_0 N_6^2$ \cite{Azeyanagi:2008kb},
and this should be determined
by the boundary theory on the D0-D6 system,
although
the corresponding Virasoro symmetries or AdS$_3$ structure
can not be directly seen from the D-brane picture 
for the KK black hole by now.
Thus, we can expect that 
the chiral CFT$_2$ of the Kerr/CFT correspondence 
for general extremal black holes
is originated in some high energy completion 
of the very near horizon geometry, 
for instance, a geometry with an AdS$_3$ throat structure.
This interpretation also 
explains why there are 
different boundary conditions corresponding to
the different chiral Virasoro symmetries 
for one very near horizon geometry.
If two or more different geometries have the same very near horizon geometry,
the appropriate boundary conditions will depend
on the original geometries.
Of course, this interpretation is not based on 
convincing evidences.
Hence it is highly important to find more convincing evidences 
for this interpretation or find 
more appropriate origin of the Kerr/CFT correspondence.

Note that we can repeat the analysis in this paper
without assuming the underlying string theory 
because the Virasoro symmetries in the CFT$_2$ dual to the AdS$_3$
can be obtained by using the Brown-Henneaux's method.
Of course, the explicit D-brane picture has been very useful to 
study the system in this paper.
In particular, the string duality will be important
for the universality of very near horizon geometries.
Although the near horizon geometry itself is changed
by taking the U-duality for the D1-D5-P case,
the string duality is expected to give examples of
the universality.

\section{Conclusions}
\label{conclusions}

In this paper, we investigated the origin of the Kerr/CFT correspondence. 
To understand this, we apply it to black holes realized as 
rotating D1-D5-P systems in string theory. 
For these black holes, we can construct different dual chiral CFT$_2$'s 
by imposing different boundary conditions on the very 
near horizon geometry.   
Geometrically, these  
black holes have an AdS$_3$ throat in the near horizon limit 
as well as a $U(1)$ fibrated AdS$_2$ geometry in the very near horizon 
limit. From this structure, for a dual chiral CFT$_2$, we found that
the Virasoro symmetry in the chiral CFT$_2$ dual to the latter originates from 
one of two Virasoro symmetries in the nonchiral CFT$_2$ dual to the former.
Since this dual nonchiral CFT$_2$ has its origin in the 
effective theory of the D1-D5 system,
we can regard the
chiral CFT$_2$ as a low energy limit,
which will leave only the ground states of the original nonchiral CFT$_2$.
Here we notice that the holographic duality is not
for one very near horizon geometry, but for the series of the
geometries with different $P$'s.

For the other dual chiral CFT$_2$'s, whose central charges are 
proportional to angular momenta,
we cannot apply such interpretations
in the similar manner as above. Instead, from the fact that 
the very near horizon geometries of extremal black holes are 
highly constrained, we expect that some different 
extremal black holes can have the same very near horizon 
geometry. In other words, there will be a kind of 
universality for such geometries. Therefore
we expect that the origin of such chiral CFT$_2$'s is 
some high energy completion of the very near horizon geometry.
For example, there might be a different extremal black hole
which has the same very near horizon and,
at the same time,  AdS$_3$ throat in the near horizon geometry.
If this is the case, the chiral CFT$_2$ is expected to have 
its origin in the nonchiral CFT$_2$ dual to the AdS$_3$ geometry.    
It will be worthwhile to find such examples.

\begin{center}
 \subsubsection*{Acknowledgements}
\end{center}

We would like to thank
Andrew Strominger for valuable comments.
We are very grateful to Keiju Murata and Tatsuma Nishioka
for explaining their results prior to publication and
for other useful discussions.
N.~O. also thanks Tohru Eguchi for helpful comments and advice.
T.~A. is supported by the Japan 
Society for the Promotion of Science (JSPS).
S.~T. is partly supported 
by the Japan Ministry of Education, Culture, Sports, Science and 
Technology (MEXT).
This work was supported by the Grant-in-Aid for the 
Global COE program ``The Next Generation of Physics, Spun from 
Universality and Emergence'' from the MEXT.

\begin{itemize}\item[]\end{itemize}
\vspace{-10mm}

\begin{center}
 \subsubsection*{Note Added}
\end{center}

As this article was being completed,
we received the preprints \cite{Chow:2008dp} and \cite{Isono:2008kx},
which partially overlap with ours.
In \cite{Chow:2008dp},
the same system as this paper was dealt with in the 5D reduced form
without the Kaluza-Klein direction,
together with many other examples.
Their results agree with ours.
In \cite{Isono:2008kx}, a special case (BMPV) is dealt with,
again in the 5D form,
although their results are different from ours and \cite{Chow:2008dp}.
They argue that the formula of central charges is modified
because of the supersymmetries.
On the other hand, our results
should be valid either with or without supersymmetries.

\vspace{10mm}


\providecommand{\href}[2]{#2}\begingroup\raggedright\endgroup

\end{document}